\documentclass[3p,times,procedia]{elsarticle}
\usepackage{nupha_ecrc}
\usepackage{subcaption}

\volume{00}

\firstpage{1}

\journalname{Nuclear Physics A}

\runauth{}

\jid{nupha}

\jnltitlelogo{Nuclear Physics A}

\usepackage{amssymb}

\usepackage[figuresright]{rotating}

\begin{document}

\begin{frontmatter}



\dochead{XXVIIIth International Conference on Ultrarelativistic Nucleus-Nucleus Collisions\\ (Quark Matter 2019)}

\title{Search for the Chiral Magnetic Effect with the ALICE detector}


\author{Sizar Aziz on behalf of the ALICE Collaboration}

\address{IJCLab, CNRS/IN2P3, Université Paris-Saclay, 91405 Orsay, France}

\begin{abstract}
In non-central heavy-ion collisions, spectator protons that do not participate in the interaction create strong magnetic fields. The strength of these fields allows testing an effect based on the hypothesized properties of QCD. The presence of so-called topological configurations can give rise to domains that carry net chirality. Coupled with the aforementioned magnetic fields, they may induce a charge separation of the particles generated in the collisions. This charge separation is called the Chiral Magnetic Effect (CME) and can be measured through charged-particle angular correlations. Measurements of the $\gamma_{1,1}$ correlator, which is sensitive to the CME, are shown for Pb--Pb collisions at $\sqrt{s_{\mathrm{NN}}} = 5.02$ TeV as well as for Xe--Xe collisions at $\sqrt{s_{\mathrm{NN}}} = 5.44$ TeV. These are found to have a significant charge dependence between opposite-sign and same-sign charge pairs. This behavior is consistent with a CME-like signal. However, the $\delta_{1}$ correlator, which measures charge correlations unrelated to any symmetry plane (i.e. background), was measured in Xe–Xe collisions and also shows a significant charge dependence. This prevents a clear interpretation of the $\gamma_{1,1}$ correlator. Novel methods to constrain the CME contribution to the $\gamma_{1,1}$ correlator are necessary.
\end{abstract}


\end{frontmatter}


\section{Introduction}
The description of Quantum Chromodynamics (QCD) includes configurations that have so-called topological charges. The QCD ground state allows for topologically non-trivial configurations, which lead to domains that carry net chirality \cite{Kharzeev:2007jp}. In the presence of strong magnetic fields, these chiral domains can induce a charge separation that is known as the Chiral Magnetic Effect (CME). In heavy-ion collisions, where the spectator protons create strong magnetic fields, it was hypothesized to be possible to measure the CME. This can be done through charged-particle angular correlations.

The relevant correlators, defined in the next section, have been measured in Au--Au and Cu--Cu collisions at a center-of-mass energy per nucleon pair of $\sqrt{s_{\mathrm{NN}}}$ = 200 GeV by the STAR Collaboration \cite{Abelev:2009ac}. At the LHC, the correlators were also studied in $\sqrt{s_{\mathrm{NN}}} = 2.76$ TeV Pb--Pb collisions \cite{Abelev:2012pa}. Although the correlators seem to indicate a signal that would correspond to the one due to the CME, the background contribution to the correlators has not yet been fully determined. 

This contribution presents the correlators measured in $\sqrt{s_{\mathrm{NN}}} = 5.02$ TeV Pb--Pb collisions and $\sqrt{s_{\mathrm{NN}}} = 5.44$ TeV Xe--Xe collisions with ALICE. By colliding different nuclei at various energies, the goal is to probe different contributions of the CME. Due to the nature of the effect, which requires strong magnetic fields, nuclei with higher Z are expected to have stronger CME contributions, and therefore stronger correlators.

\section{Two- and three-particle correlators}
The original correlator proposed to measure the CME was the so-called $\gamma_{1,1}$ correlator, defined as 
\begin{equation}
    \gamma_{1,1} = \langle \cos(\varphi_{\alpha} + \varphi_{\beta} - 2 \Psi_\mathrm{RP}) \rangle.
    \label{eq:gamma11}
\end{equation}
Here, $\varphi$ represents the azimuthal angle of a track while $\alpha$ and $\beta$ indicate its charge. One can, therefore, either consider pairs of tracks with the same or opposite sign. The reaction plane angle $\Psi_{\mathrm{RP}}$ is defined by the impact parameter and beam directions. The angled brackets indicate that the average is taken over all possible particle pairs in an event. The magnetic field is generated perpendicular to the reaction plane. In addition, the separation due to the CME will occur relative to the reaction plane. However, the reaction plane is experimentally inaccessible and must be approximated by one that can be obtained from the tracks. The second order event plane $\Psi_{2}$ is therefore used. It arises from the fourier decomposition of the azimuthal distributions and is correlated with elliptic flow $v_{2}$ \cite{Poskanzer:1998yz}. In addition, two particle correlators not referring to any symmetry plane can be defined. The one used in this analysis is the $\delta_{1}$ correlator defined as 

\begin{equation}
    \delta_{1} = \langle \cos(\varphi_{\alpha} - \varphi_{\beta}) \rangle,
\end{equation}
where $\delta_{1}$ is also calculated for both opposite and same sign charge pairs. By construction, $\delta_{1}$ mainly probes correlations unrelated to any symmetry plane. It is therefore dominated by so-called non-flow effects, which are correlations that do not arise due to the initial geometry of the collisions. These non-flow effects include processes such as resonance decays, jets and transverse momentum conservation. The size of $\delta_{1}$ gives an indication on the extent to which non-flow is present in the correlator, i.e. an estimate of the background contribution in the $\gamma_{1,1}$ correlator.

\section{Event and track selections}
All events were recorded by the central barrel of the ALICE detector \cite{Aamodt:2008zz}. Results from two analyses are presented here: 5.02 TeV Pb--Pb collisions as well as 5.44 TeV Xe--Xe collisions. In both cases, minimum bias events were used in the centrality interval between 0 and 90\%. Approximately 1.3 and 40 million of such events were used for the Xe–Xe and Pb–Pb analyses, respectively. The tracks considered for analyses are primary tracks with a pseudorapidity $|\eta| < 0.8$, matching the acceptance of the ALICE central barrel detectors. In addition, a selection on the transverse momentum $0.2 < p_{\mathrm{T}} < 5.0$ GeV/$c$ is applied.

\section{Results}
\begin{figure}[h!]
    \centering
    \includegraphics[height=5cm]{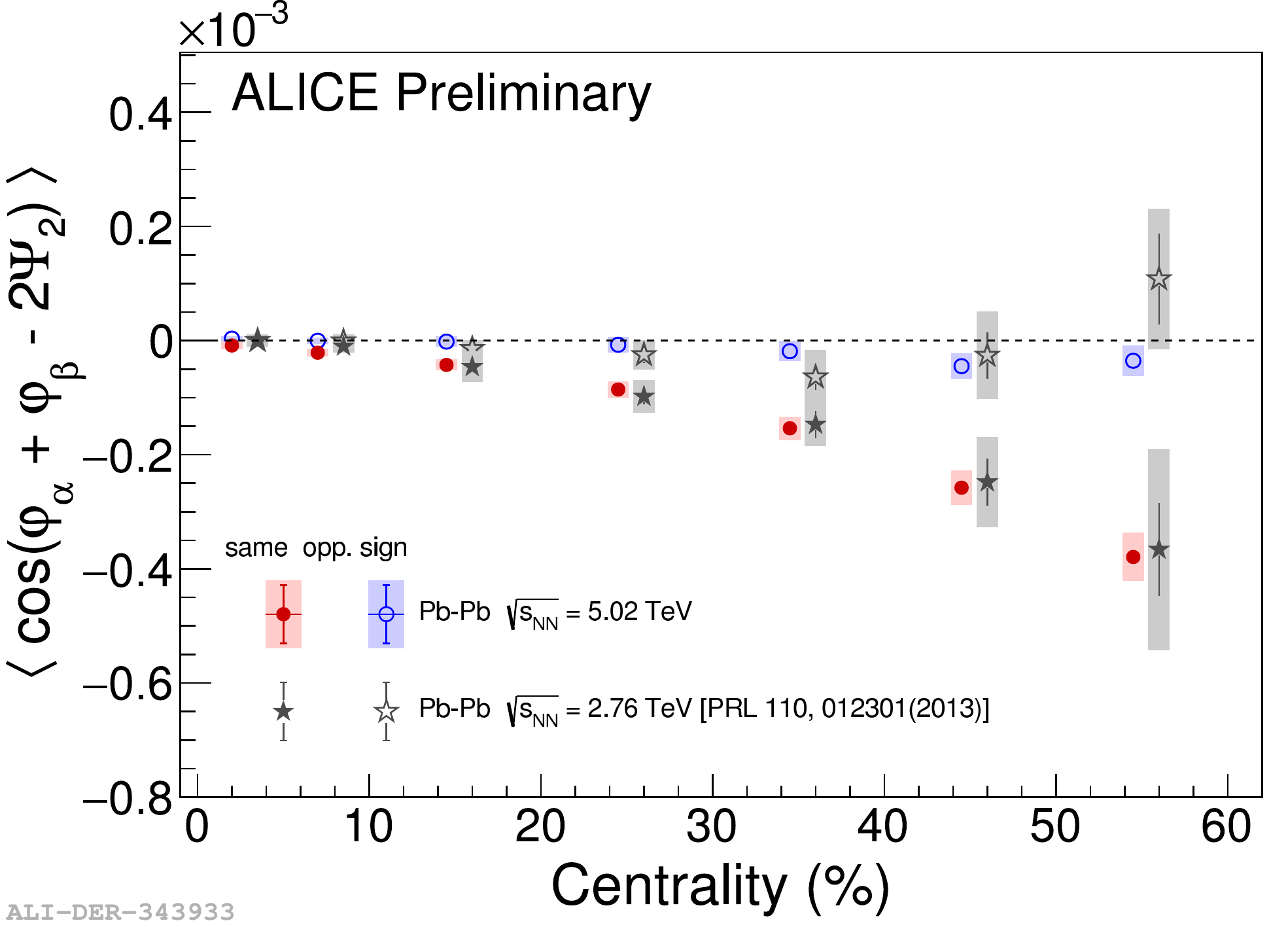}
    \includegraphics[height=5cm]{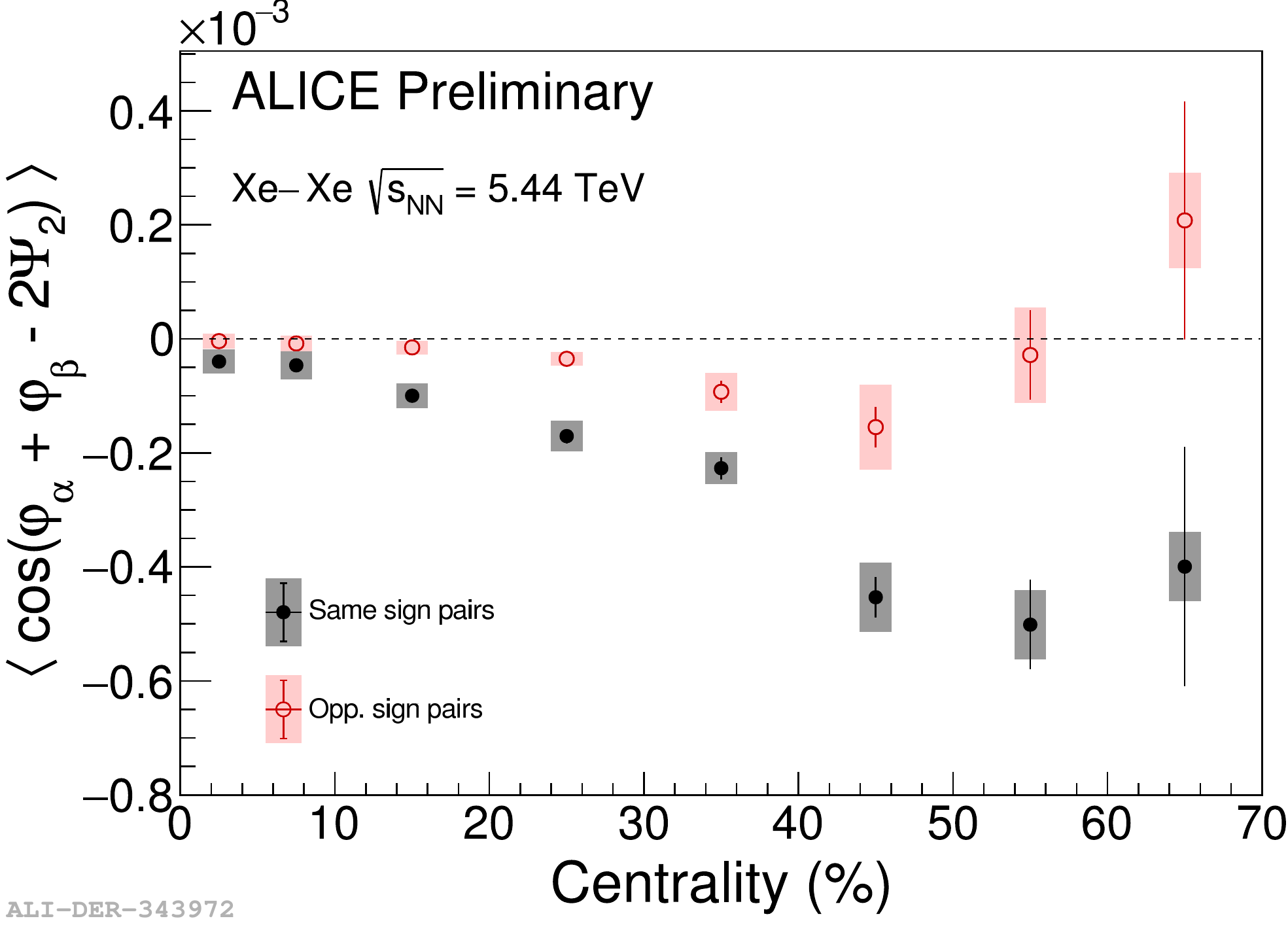}
    \caption{Centrality distribution of the $\gamma_{1,1}$ correlators for both same-sign and opposite-sign charge pairs. On the left hand side are the correlators for Pb--Pb collisions at $\sqrt{s_{\mathrm{NN}}} = 2.76$ as well as 5.02 TeV. On the right hand side are the results for $\sqrt{s_{\mathrm{NN}}} = 5.44$ TeV Xe--Xe collisions.}
    \label{fig:gammaCor}
\end{figure}

Figure~\ref{fig:gammaCor} shows the $\gamma_{1,1}$, for Pb–Pb (left) and Xe–Xe (right) collisions. 
In addition to new results from Pb–Pb collisions at $\sqrt{s_{\mathrm{NN}}} = 5.02$ TeV, previously published ALICE results at 2.76 TeV are shown as well \cite{Abelev:2012pa}. Across different collision energies and systems, a clear charge dependence is seen for the pairs with the same and oppositely charged particles. For opposite-sign pairs, values are close to 0 and show a weak centrality dependence. The correlator for same-sign pairs is also very small in central collisions. However, it becomes increasingly more negative towards peripheral collisions. Therefore, the difference grows. This behaviour is seen in Figure~\ref{fig:gammaCor} for both Pb--Pb (left) as well as Xe--Xe (right). It is consistent with a CME-like signal. Central collisions by nature contain few spectator nucleons, leading to weaker magnetic fields which in turn prevents the CME from manifesting.

\begin{figure}[h!]
    \centering
    {\includegraphics[height=5cm]{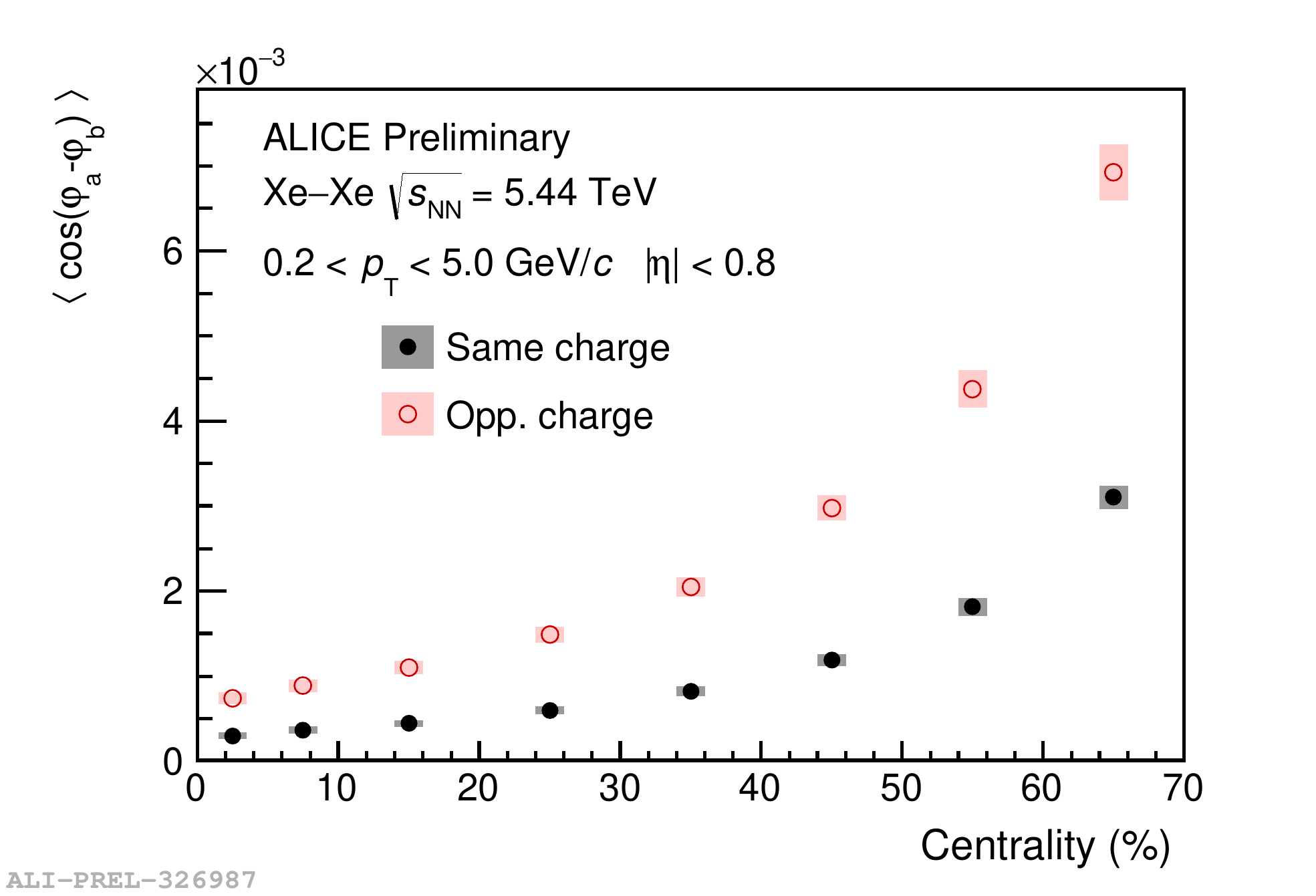}}
    {\includegraphics[height=5cm]{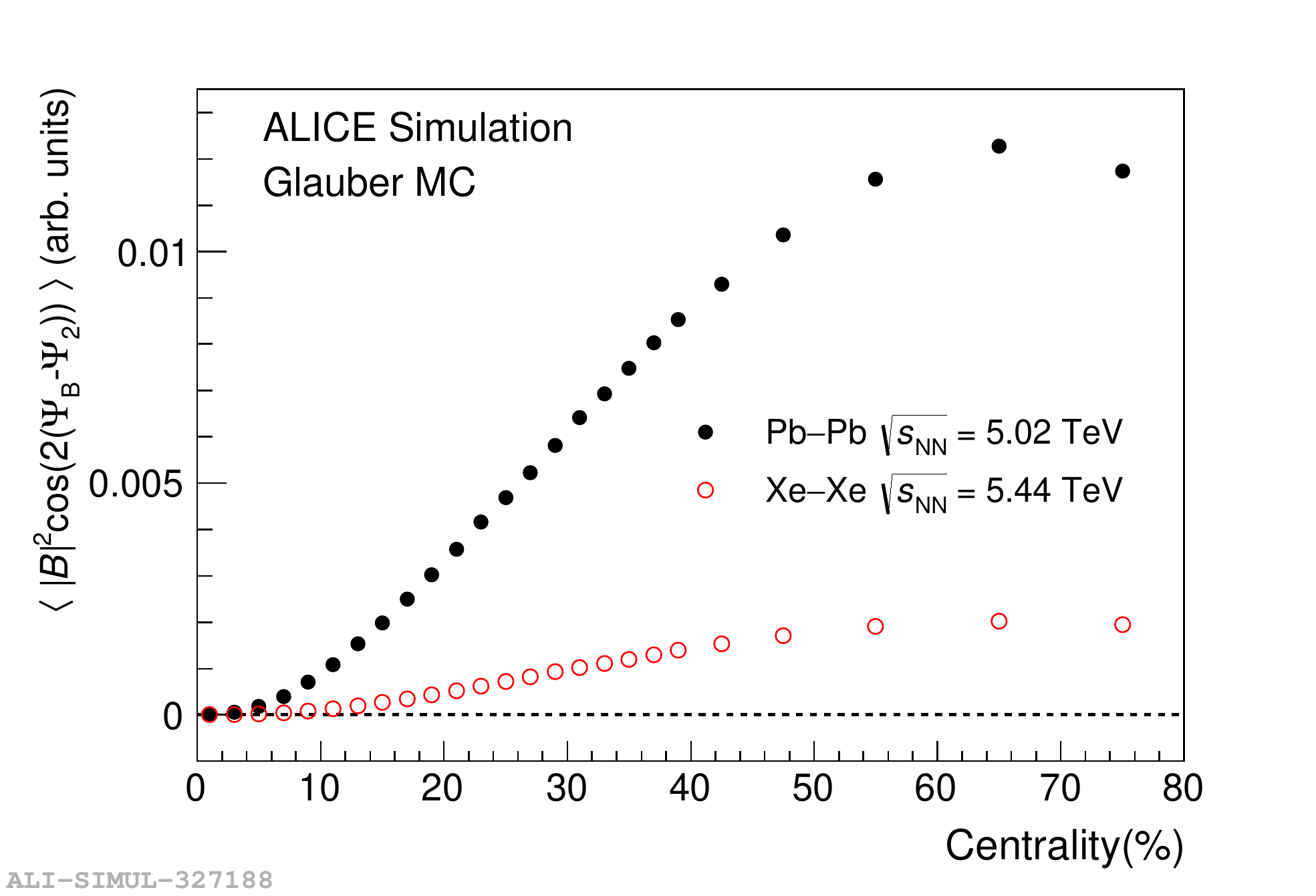}}
    \caption{(left) $\delta_{1}$ correlator for Xe--Xe collisions, which measures charge correlations unrelated to any symmetry plane. (right) Glauber Monte Carlo simulation of magnetic field strength comparison for Pb--Pb versus Xe--Xe collisions as a function of centrality.}
    \label{fig:deltaCor+magfield}
\end{figure}

The left panel of Fig.~\ref{fig:deltaCor+magfield} shows the $\delta_{1}$ correlator for Xe--Xe collisions. The same trend is seen as for $\gamma_{1,1}$, i.e. a difference between opposite-sign and same-sign pairs that grows towards peripheral collisions. This complicates the interpretation of the signal in $\gamma_{1,1}$ as a significant amount of the measured correlator could be due to background. Unfortunately, there is no direct way to relate the $\delta_{1}$ correlator to a background fraction in $\gamma_{1,1}$. The right panel of Fig.~\ref{fig:deltaCor+magfield} shows a Monte Carlo Glauber simulation of the magnetic field strength in the collisions, calculated as described in Reference \cite{Kharzeev:2007jp}. The values are given in the center of the collision at a time of $\tau = 0.1 \mathrm{fm}/c$. The magnetic field is several times stronger for Pb--Pb collisions than for Xe--Xe collisions. While that can be expected due to the difference in proton number $Z$, this presents a puzzle. In Fig.~\ref{fig:gammaCor}, the separate opposite-sign and same-sign correlators as well as their difference are quantitatively close to each other for Xe--Xe and Pb--Pb collisions. However, a stronger magnetic field should separate the charges more strongly, increasing (at least) the difference of the measured correlators. This is interpreted as another indication that background processes could be dominating the measurement. Studies of the background have already been done by both ALICE and CMS collaborations. Based on the event shape engineering method, an upper limit on the CME contribution of 26-33\% with a confidence level of 95\% was derived in ALICE \cite{Acharya:2017fau}. Using extrapolations from the $\gamma_{1,1}$ measured in p--Pb collisions, the CMS study, however, concluded an upper limit of 7\% at 95\% confidence level \cite{Sirunyan:2017quh}. It has become clear throughout the years that the CME might only be responsible for a small fraction of the measured $\gamma_{1,1}$ correlator. However, as seen in the different studies, it remains to be determined what the exact size is.

\begin{figure}[h!]
    \includegraphics[height=5cm]{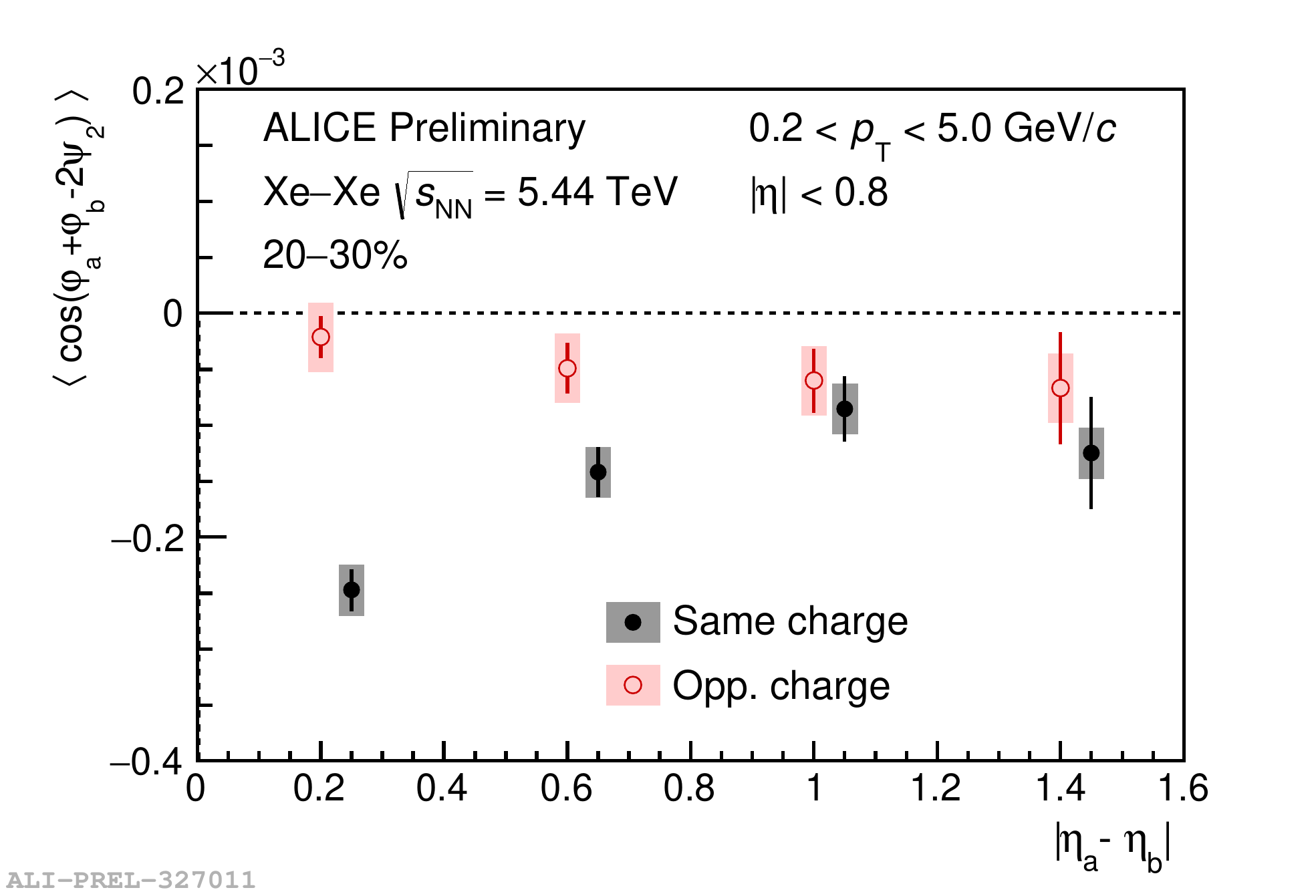}
    \includegraphics[height=5cm]{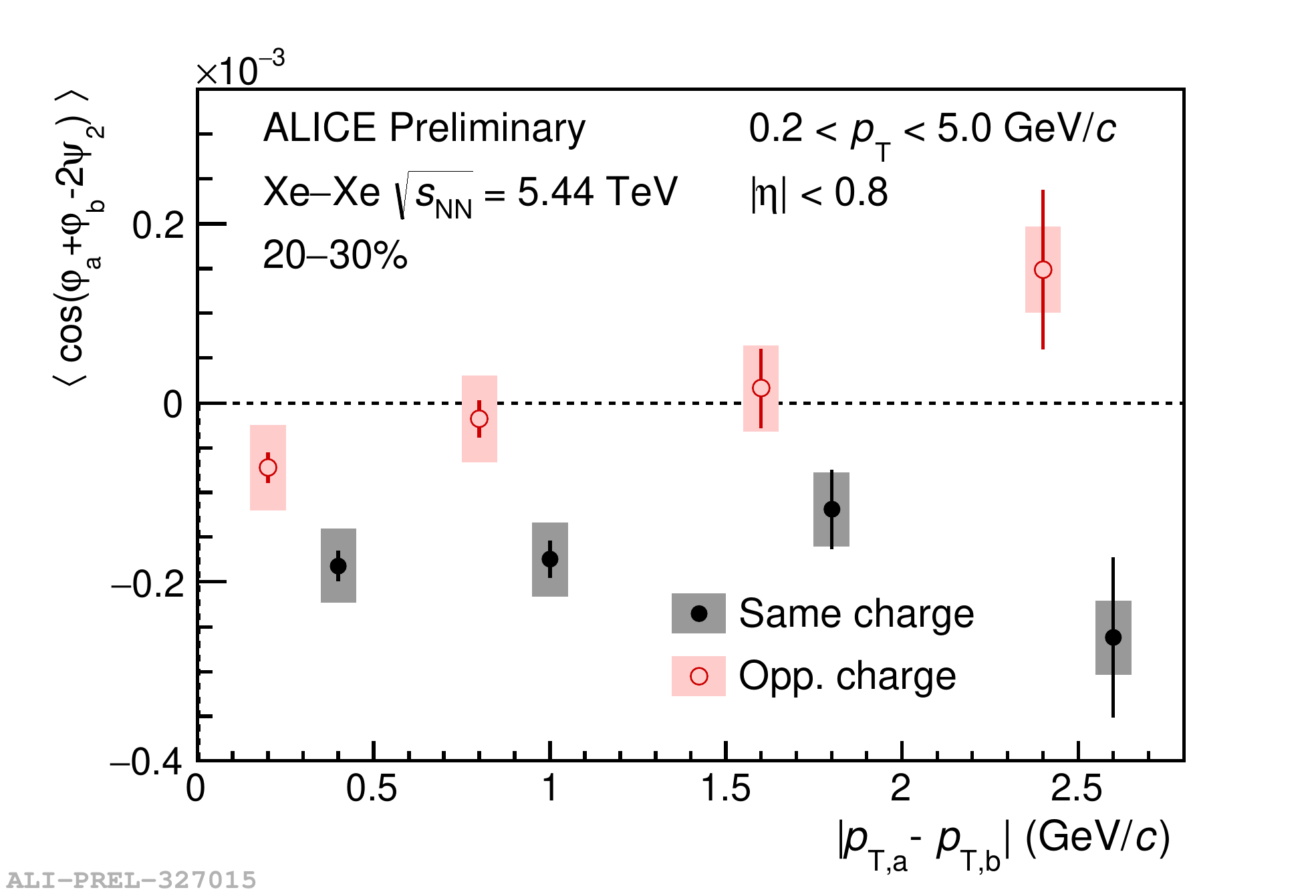}
    \caption{Differential results of $\gamma_{1,1}$ for Xe--Xe collisions. The left plot shows the correlator as a function of the $\eta$ difference (left) and $p_{\mathrm{T}}$ difference (right) of the two tracks.}
    \label{fig:diffGamma}
\end{figure}

In order to probe whether a specific kinematic region particularly contributes to $\gamma_{1,1}$, a more differential analysis was done. $\gamma_{1,1}$ was studied as a function of the pseudorapidity difference $\eta_{\alpha} - \eta_{\beta}$, $p_{\mathrm{T}}$ difference and average $p_{\mathrm{T}}$. Two different patterns are observed. For the $\eta$ difference and average $p_{\mathrm{T}}$, there is a weak dependence for the opposite-sign pairs, while there is a stronger dependence for same-sign pairs. This is seen on the left panel of Fig.~\ref{fig:diffGamma}. The $\gamma_{1,1}$ correlator stays roughly constant for opposite-sign pairs while the values for same-sign pairs depend slightly on the $\eta$ difference. The opposite behavior is seen in for the $p_{\mathrm{T}}$ difference, on the right panel of Fig.~\ref{fig:diffGamma}. 

\section{Conclusion and outlook}
The $\gamma_{1,1}$ correlator was measured in Pb--Pb collisions at $\sqrt{s_{\mathrm{NN}}} = 5.02$ TeV and Xe--Xe collisions at $\sqrt{s_{\mathrm{NN}}} = 5.44$ TeV. In addition, for the Xe--Xe collisions, $\delta_{1}$ was measured as well. In both collision systems, a strong charge dependence is seen in $\gamma_{1,1}$, which is consistent with a CME-like signal. However, the $\delta_{1}$ correlator, which measures background contributions unrelated to the symmetry plane, also shows a significant charge dependence. This is an indication that a significant percentage of the $\gamma_{1,1}$ correlator is actually due to background. This is supported by the similarity of the $\gamma_{1,1}$ correlator in Pb--Pb and Xe--Xe collisions even though the strength of the magnetic field is much larger in the former. Studies are necessary to constrain the possible CME contribution to the $\gamma_{1,1}$ correlator quantitatively. A new method to estimate the CME contribution is currently in development, correlating different event planes. Under certain assumptions, it could provide another approach to estimate the upper limit on the possible CME contribution.





\bibliographystyle{elsarticle-num}
\bibliography{refs.bib}







\end{document}